\documentclass[11pt,twoside]{article}
\usepackage{asp2010}
\resetcounters

\begin{document}

\title{Cloud Computing with Context Cameras}
\author{A.J. Pickles and W.E. Rosing
\affil{Las Cumbres Observatory Global Telescope,\\
6740 Cortona Drive, Goleta CA 93117}}

\begin{abstract}

We summarize methods and plans to monitor and calibrate photometric
 observations with our autonomous, robotic network of 2m, 1m and 40cm
 telescopes. These are sited globally to optimize our ability to observe
 time-variable sources. 
 
Wide field ``context'' cameras are aligned with our network telescopes
 and cycle every $\sim$2 minutes through BVr'i'z' filters, spanning our
 optical range.  We measure instantaneous zero-point offsets and
 transparency (throughput) against calibrators in the 5-12m range from
 the all-sky Tycho2 catalog, and periodically against primary standards.
Similar measurements are made for all our science images, with typical
 fields of view of $\sim$0.5 degrees. These are matched against Landolt,
 Stetson and Sloan standards, and against calibrators in the 10-17m
 range from the all-sky APASS catalog. Such measurements provide pretty
 good instantaneous flux calibration, often to better than 5\%, even in
 cloudy conditions.

Zero-point and transparency measurements can be used to characterize,
 monitor and inter-compare sites and equipment. When accurate
 calibrations of Target against Standard fields are required, 
 monitoring measurements can be used to select truly photometric periods
 when accurate calibrations can be automatically scheduled and
 performed.

\end{abstract}

\section{Introduction}

Las Cumbres Observatory Global Telescope (LCOGT) is equipping a global
network of 2m, 1m and 40cm telescopes with homogeneous
instrumentation. Our goal is to provide maximally available monitoring
of time variable sources, from solar system to extra-galactic objects,
and ranging in brightness from about 7-20m.
Ideally we would like to provide accurate relative light curves with
accurate absolute photometric calibration for all our imaging data, but the
latter would preclude observations in non-photometric conditions, and
would require a large overhead of standard observations that would take
time away from the many programs that do not require such absolute accuracy.

Many observations of time-variable sources are able to achieve very high
relative photometric precision, typically to a few mmag, by comparison
with stars within the target field. Observations of exoplanet transits
for example do not require absolute flux calibration, and in many cases
results are quite insensitive to variable atmospheric transmission, or
filter passband.  Similar statements may be made for other time variable
observations such as micro-lensing events, but in this case data from
several sites, instruments and passbands are likely to be combined, so
reasonable flux calibration of the different data sets enables more
reliable combinations and comparisons.

More demanding applications include SN light curve observations, where
accurate flux calibration to 1\% or better is required, to minimize the
effects of observational error on derived cosmological parameters. These
accurate calibrations can be performed {\em post-facto} on target
fields, after light curves of interesting targets have been determined
with good relative accuracy. A number of target fields requiring
accurate flux calibration can be combined into a single calibration
program during photometric periods.

We are therefore trending towards a two-stage approach for our
photometric calibrations. The first stage includes automatically
measuring a zero-point magnitude offset to better than 10\% (goal
$\le$5\%) for all images, based on available calibrators within each
observed field. Measured zero points and rms errors are attached to each
FITS header, and can be added to instrumental mags (eg. as
MAG\_ZEROPOINT in {\em
sextractor}\footnote{http://www.astromatic.net/software/sextractor}) to
provide calibrated magnitudes to within the recorded error.

The second stage, for more accurate photometric calibration but of fewer
targets, requires the observer to propose specific targets and
methodology to derive detailed calibrations for these target fields. In
the latter case, LCOGT can provide the continuous monitoring to indicate
when photometric conditions prevail, and when these demanding
calibrations should be automatically scheduled.

\section{All-Sky Catalogs}

It has become common practice to provide rapid and automatic World
Coordinate System (WCS) fits to astronomical images, usually by
reference to the USNO-B or NOMAD catalogs \citep{zacharias_2004}. We
perform WCS fits with {\em astrometry.net}\footnote{http://astrometry.net}
for instantaneous or ``flash'' reductions, and with {\em wcsfit} within the
ORAC-DR pipeline \citep{bridger_1998} for final data products.
With the advent of comprehensive all-sky photometric catalogs, it is a
short step to add pretty good automatic photometric calibration. Our
experience is that this can be accurate to a few percent, even in quite
cloudy observing conditions.

The Tycho2 catalog \citep{hog_2000} contains $\sim$2.5M stars around the
sky, with an average density of $\sim$60 stars per square degree, and a
useful magnitude range from 3-12mag. It is well suited to wide-format
devices such as SkyProbe \citep{steinbring_2009}, and the 3-degree wide
Context cameras described here. The Tycho2 catalog provides native
passbands in B$_T$ and V$_T$. Cousins R-mags can be added from NOMAD
\citep{zacharias_2004} as can infrared JHK$_S$ bands from the 2MASS
survey \citep{cutri_1998}.  Fitted magnitudes in any desired system
passbands can be obtained by matching a best-fit library spectrum to the
catalog passbands. A version of automatic flux calibration utilizing
Tycho2 was proposed in \citet{pd_2010}.

The number and magnitude range of the Tycho2 catalog provide relatively
sparse automatic flux calibration within typical cassegrain images
covering less than one degree, and their magnitude range is too bright
for many telescopes. Large survey cameras such as PTF in the Northern
hemisphere can automatically calibrate against extensive Sloan survey data
\citep{ofek_2012}. In a few years the Gaia mission (E. Pancino, this
conference) promises to provide a dense all-sky astrometric and
photometric catalog to satisfy different telescope apertures, and image
fields.

Fortunately the AAVSO Photometric All-Sky Survey
(APASS\footnote{http://www.aavso.org/apass}, A. Henden, this conference)
now provides over 40M stars across most of the sky. APASS DR6 provides
good flux calibrators at a typical density of over one thousand stars
per square degree, in the 10-17mag range, and accurate to about 3\% in
passbands BVg'r'i'. Spectral matching to BVg'r'i' and corresponding
2MASS JHK$_S$ magnitudes, again enables synthetic mags to be calculated
for other desired passbands.  We have started using the APASS DR6
catalog on our science images and find excellent results. We typically
find $\ge$50 matching stars in our current 15x15 arcmin frames (to be
increased to 27x27 arcmin), resulting in a measured zero-point rms (for
stars of varying brightness and color) in the range 3--5\% for
un-flatfielded observations, and often better than 3\% for flat-fielded
data, in atmospheric conditions varying from clear to quite cloudy.

\section{Extinction and Transparency}

Here we use the term ``Extinction'' to refer to clear atmospheric
extinction that is measured over a reasonable time ($\ge$1-night) to
vary linearly with airmass. Extinction caused by Rayleigh scattering,
ozone, and aerosols in an atmospheric shell around the Earth is then
considered to be well behaved, or ``photometric'' for that period, with
coefficients of extinction per airmass (per passband) that are well
defined and measurable. Extinction calibrations may also include color
terms to match observed to standard passbands, and may include terms to
allow for variable atmospheric absorption or emission features that
affect one or more passbands.

The term ``Transparency'' here quantifies any conditions, including
cloudy weather or conditions of significant non-photometric atmospheric
variability. Transparency varies from a maximum of 100\% (photometric if
maintained over a significant period) to much lower values, and becomes
unmeasurable below about 10\%. Useful observations can still be obtained
when the transparency is as low as 20--30\% however. Transparency
calculations here allow for the nominal clear-atmosphere extinction per
passband and airmass, where the extinction coefficients can be calculated from
\citet{hayes_1975}, or calibrated by on-sky measurements.

\section{Telescope Throughput}

\begin{figure}[!ht]
\plotone{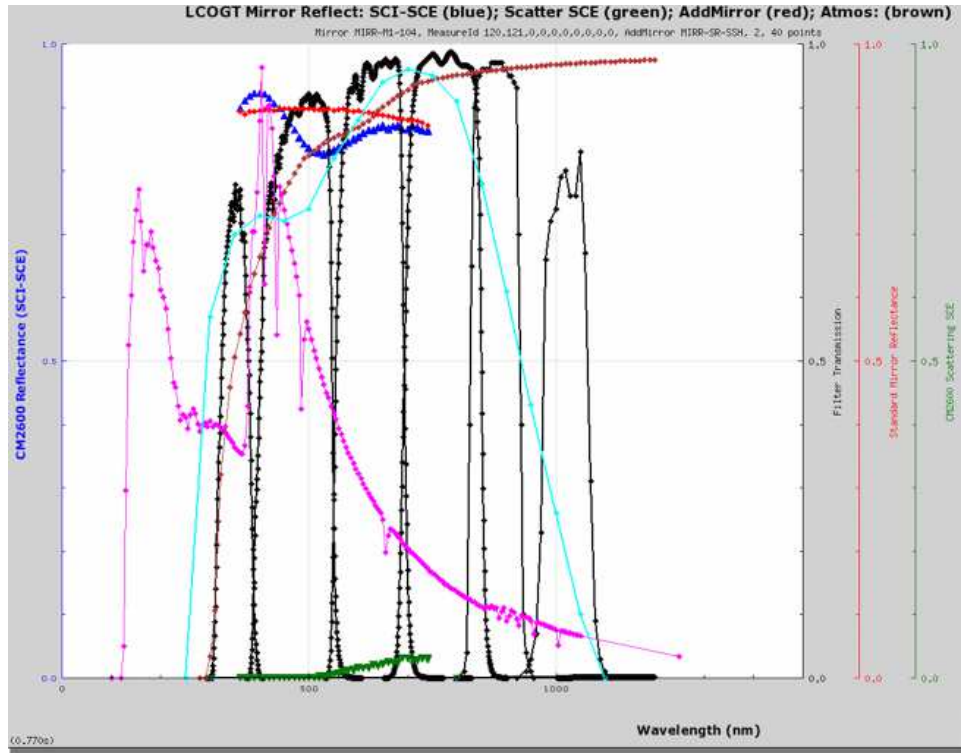}
\caption{Throughput measurements from our database.
\label{fig:throughput}
}
\end{figure}

Telescope optical throughput includes reflectivities of mirrors,
transmission of corrective optics and filters, and QE of
detectors, all of which can vary (slowly) with time. Figure
\ref{fig:throughput} is a screenshot showing the attenuation of an A0~V
spectrum (magenta) by various components extracted from our database:
mirrors (standard reflectance curve in red, measured single-mirror
reflectance from 380--760~nm in blue, scattering in green), a smooth atmospheric
transmssion (brown) calculated from \citet{hayes_1975} for 1.3 airmasses
at an elevation of 2200m, filter transmissions (black) and CCD QE
(cyan).

\section{Filters and System Passbands}

We have designed and ordered custom filters from Astrodon, to optimally
match UBVRI standards \citep{landolt_2009}, Landolt (this conference)
and
Stetson\footnote{http://www3.cadc-ccda.hia-iha.nrc-cnrc.gc.ca/community/STETSON/standards/}. We
were able to match our system UBVRI passbands to those of
\citet{bessell_1979, bessell_1991, maiz_2006} that provide the best
matches to Landolt standards (with minimized color corrections) when
convolved with calspec flux-calibrated spectra \citep{pickles_2010}.

\begin{figure}[!ht]
\plottwo{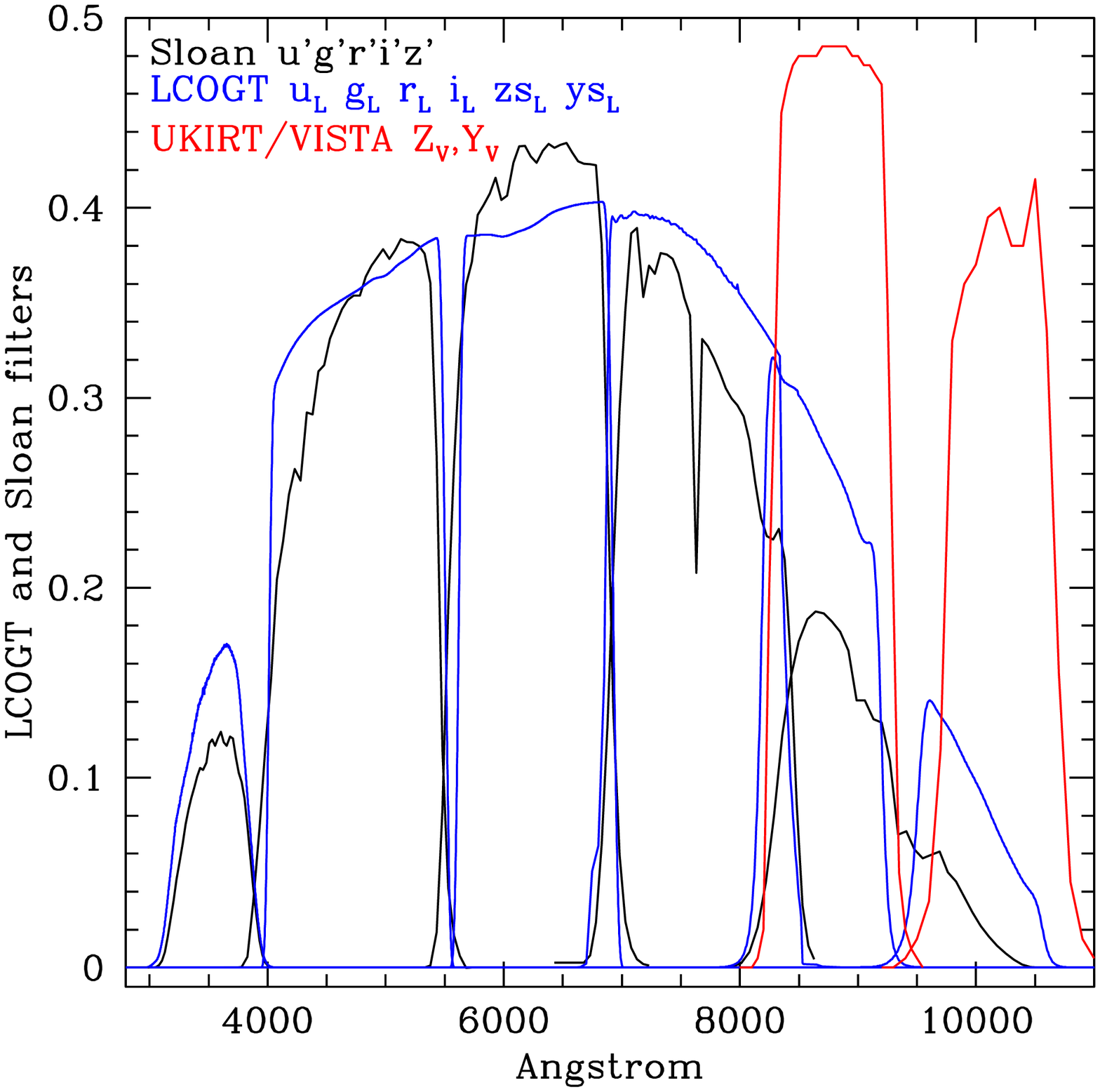}{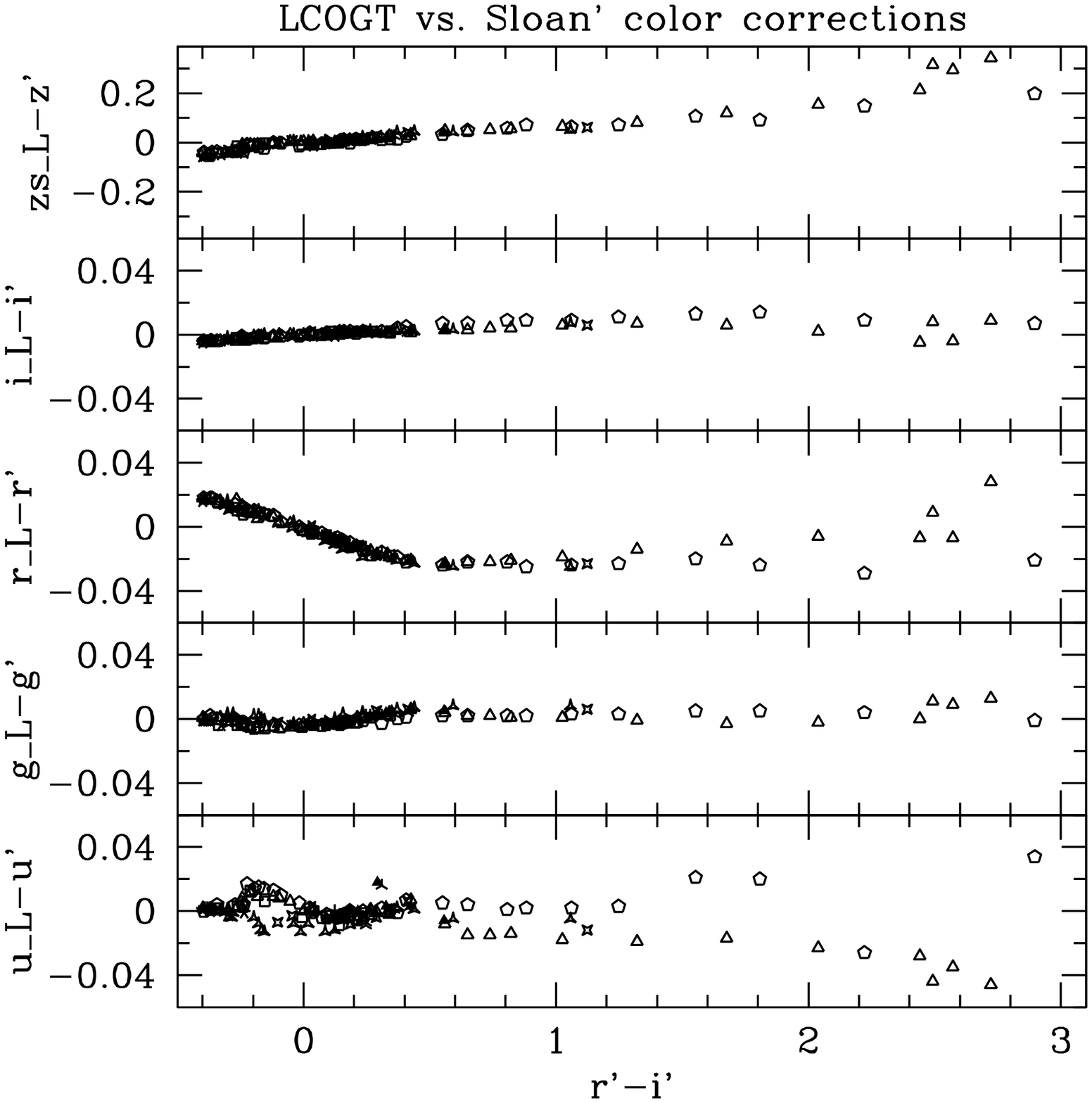}
\caption{LCOGT filter passbands and modeled color corrections
\label{fig:filters}
}
\end{figure}

We have adjusted our Sloan primed passbands slightly from those listed
for the USNO 40-in
telescope\footnote{http://www-star.fnal.gov/ugriz/Filters/response.html}.
Our g$_L$ filter has its blue edge shifted redwards to 405nm. Our g$_L$
red edge and r$_L$ blue edge were optimized to exclude the night-sky OI
5577~\AA\ feature. Our zs$_L$, ys$_L$ filters are chosen to match the
short red-cutoff Pan-Starrs z$_S$, y$_S$
filters\footnote{http://svn.pan-starrs.ifa.hawaii.edu/trac/ipp/wiki/PS1\_Photometric\_System}. Our
w$_L$ filter matches the wide combination of g+r+i.  LCOGT ugrizy$_L$ system
passbands (smooth atmosphere, telescope, filter, detector) are shown on
the left of Figure \ref{fig:filters} for u$_l$ g$_L$, r$_L$, i$_L$
zs$_L$, ys$_L$ in blue, compared with Sloan USNO-40 primed filters
(black) and UKIRT/VISTA Z$_V$,Y$_V$ (filter only) in red.

It is not possible to match observed to standard system passbands
precisely, because of variations in telescope throughput, atmosphere and
detector QE. It is possible to model (and check by measurement) the
color differences that result.  We have convolved LCOGT and Sloan primed
system passbands against a set of standard spectra and computed the
color difference as a function of color. These are shown on the right in
Figure \ref{fig:filters}. For detailed on-sky comparisons, we compare
against Sloan standards from \citet{smith_2002, smith_2005, tucker_2006,
davenport_2007}.

\section{Flat-fielding and Baffling}

Removal of the instrument signature should remove pixel-to-pixel
variations, zonal effects including vignetting, and variable QE response
to uniform illumination. After signature removal a star should give the
same flux and instrumental magnitude anywhere in the detector field. But
errors in the flat-field system, illumination and color differences
between the sky illumination and flat-field, and scattered (stray) light
from imperfect baffling can all lead to residual response variations in
the 1--3\% range across the detector. These effects need to be
understood, minimized and calibrated. We were able to reduce this effect
significantly on our 2m telescopes by improved light baffling
\citetext{Rosing and Tufts, in preparation}. We have devoted
considerable attention to designing good 1m telescope baffles, and
minimizing ghosting \citep{haldeman_2010}, and to designing a uniform
flat-field illumination system for all our telescopes
\citep{Haldeman_2008}.

\section{Context Camera and Results}

\begin{figure}[ht!]
\plottwo{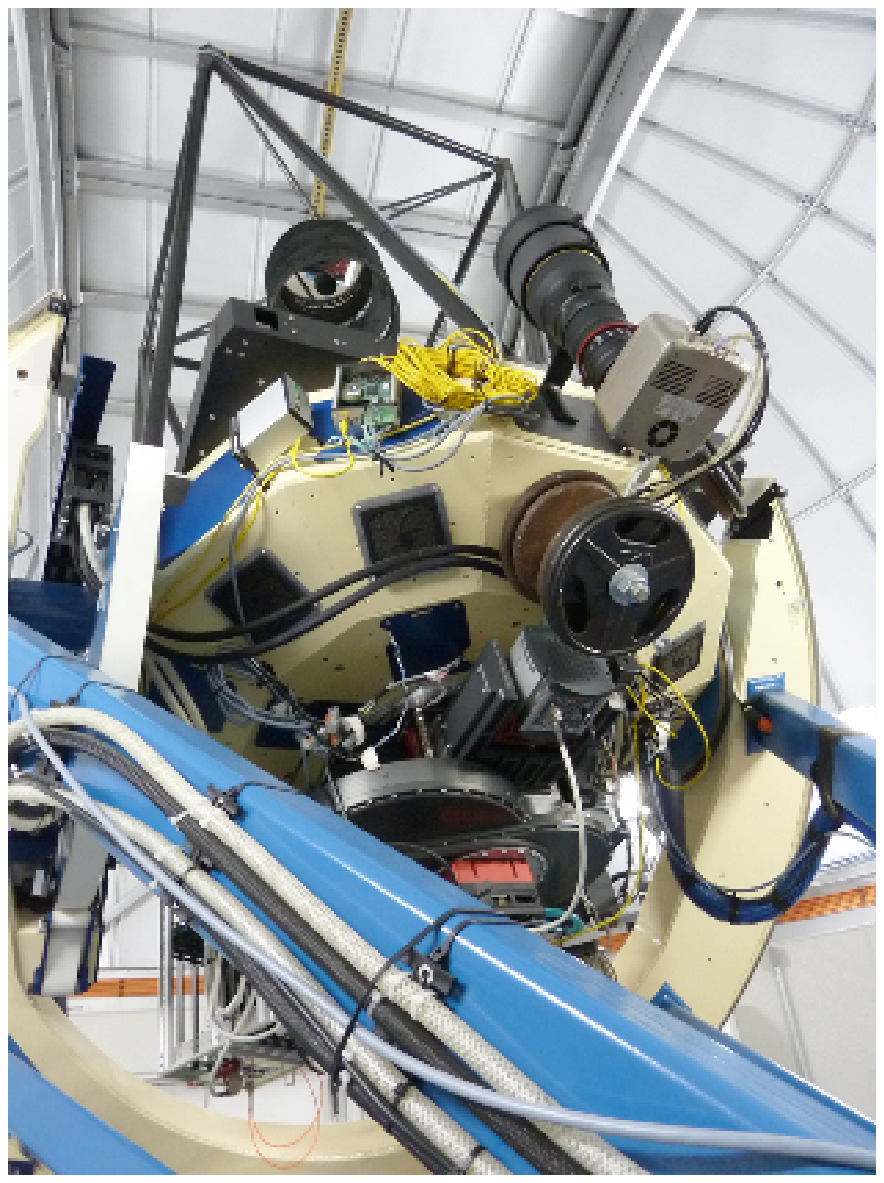}{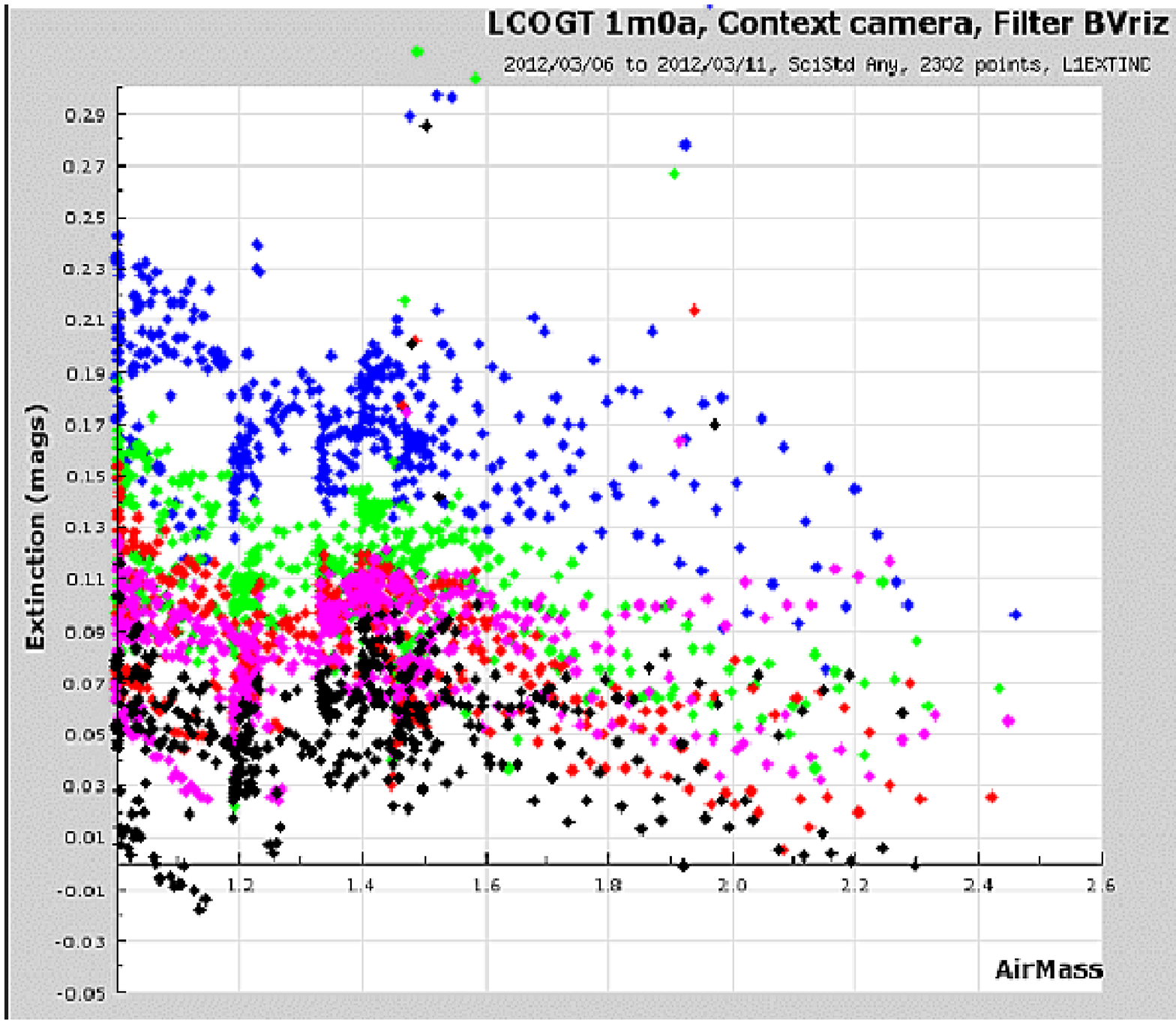}
\caption{Context camera mounted on our 1m telescope in Santa Barbara and extinction measurements vs. airmass
\label{fig:con_cam}
}
\end{figure}

The context camera results illustrated here use a Nikon 400mm f/2.8 lens
with 5-position filter wheel and SBIG 6803 CCD.  This results
in 3072x2048 images with 4.7-arcsec pixels, and a 4x2.7 degree field of
view. Figure \ref{fig:con_cam} shows such a device
mounted on the south side of a 1m telescope mirror cell, coaligned
with the 1m telescope pointing. We are also testing other 
camera setups, on independent mounts, that can service multiple pointings
\footnote{http://lcogt.net/network/0.4m}.

The context camera cycles through its 5 filters (BVr'i'z') in about
2-minutes. Each frame is WCS fitted, calibrating stars within the
pointing area\footnote{http://lcogt.net/ajp/findstars} are coordinate
matched with observed stars in each image to within a small tolerance
($\sim$3-arcsec), usually resulting in $\ge$100 matches.  For each image
the differences between {\em sextractor} instrumental and catalog
magnitudes in the relevant passband are calculated, and a mean and rms
formed. Because of vignetting in the system, the un-flat-fielded context
data typically have about 15\% rms, but flat-fielded or simply
un-vignetted images typically have $\le$5\% rms. The highest values of
these differences correspond to clear, possibly photometric
weather; lower values to cloudy conditions. Measurements are
displayed in real-time on the system, and stored in a database.

An example of these results for a 6-night period in March from our
facility in Santa Barbara is shown as a screenshot from our database to
the right in Figure \ref{fig:con_cam}. Extinction in magnitudes is
plotted against airmass for B--blue, V--green, r'--red, i'--magenta and
z'--black. Measured extinctions show trends with
airmass in each color, but none of these data indicate photometric
conditions. Photometric nights are rare in Santa Barbara, particularly
at our facilites close to UCSB, at low elevation, and close to the sea.

We have installed a Context camera at our recently deployed 1-m at
McDonald Observatory, and will extend them with 1m telescope deployments
this year to our sites at CTIO (Chile), SAAO (South Africa) and SSO
(Australia). Our goal is to accumulate data and calibrations with a view to
characterizing each site, and automatically detecting photometric
periods.

\section{Conclusion}

Wide-field Context cameras are well suited to monitoring the changing
atmosphere at astronomical sites. They provide quick, automatic and reliable
calibrations of transparency. More importantly for our purposes, they
can automatically detect and signal photometric conditions when they
prevail. At these times detailed calibrations of extinction
coefficients, and of important target fields, can be scheduled with
network telescopes.

\acknowledgements We are happy to acknowledge the American Association
of Variable Star observers (AAVSO) and their Robert Martin Ayers Science
funding, for providing a comprehensive all-sky catalog. We thank the
organizers of this conference for providing the opportunity to learn
more about methods for characterizing and measuring atmospheric variability.

\bibliography{pickles_a}

\begin{thebibliography}{}
\expandafter\ifx\csname natexlab\endcsname\relax\def\natexlab#1{#1}\fi
\expandafter\ifx\csname url\endcsname\relax
  \def\url#1{\texttt{#1}}\fi
\expandafter\ifx\csname urlprefix\endcsname\relax\def\urlprefix{URL }\fi
\providecommand{\eprint}[2][]{\url{#2}}

\bibitem[{Bessell(1979)}]{bessell_1979}
Bessell, M. 1979, PASP, 91, 589

\bibitem[{Bessell(1991)}]{bessell_1991}
--- 1991, AJ, 101, 662

\bibitem[{Bridger et~al.(1998)Bridger, Economu, Wright, \&
  Currie}]{bridger_1998}
Bridger, A., Economu, F., Wright, G., \& Currie, M. 1998, proc. SPIE, 3349, 184

\bibitem[{Cutri(1998)}]{cutri_1998}
Cutri, R. 1998, AAS, 192, 6402

\bibitem[{Davenport et~al.(2007)Davenport, Bochanski, Covey, Hawley, West, \&
  Scheider}]{davenport_2007}
Davenport, J., Bochanski, J., Covey, K., Hawley, S., West, A., \& Scheider, D.
  2007, AJ, 134, 2430

\bibitem[{Haldeman et~al.(2010)Haldeman, Hayes, Posner, Tufts, Pickles, \&
  Dubberley}]{haldeman_2010}
Haldeman, B., Hayes, R., Posner, V., Tufts, J., Pickles, A., \& Dubberley, M.
  2010, proc. SPIE, 7739, 56

\bibitem[{Haldeman et~al.(2008)Haldeman, Tufts, Hidas, Dubberley, \&
  Posner}]{Haldeman_2008}
Haldeman, B., Tufts, J., Hidas, M., Dubberley, M., \& Posner, V. 2008, proc.
  SPIE, 7014, 67

\bibitem[{Hayes \& Latham(1975)}]{hayes_1975}
Hayes, D., \& Latham, D. 1975, ApJ, 197, 593

\bibitem[{H$\o$g et~al.(2000)H$\o$g, Fabricius, Makarov, Urban, Corbin, G.,
  Bastian, Schwekendick, \& A.}]{hog_2000}
H$\o$g, E., Fabricius, C., Makarov, V., Urban, S., Corbin, T., G., W., Bastian,
  U., Schwekendick, P., \& A., W. 2000, AA, 355, 27

\bibitem[{Landolt(2009)}]{landolt_2009}
Landolt, A. 2009, AJ, 137, 4186

\bibitem[{Maiz~Apellaniz(2006)}]{maiz_2006}
Maiz~Apellaniz, J. 2006, AJ, 131, 1184

\bibitem[{Ofek et~al.(2012)Ofek, Laher, Surace, D., Sesar, Horesh, Law, van
  Eyken, Kulkarni, Prince, Nugent, Sullivan, Yaron, Pickles, Agueros, Bildsten,
  Cenko, Gal-Yam, Grillmair, Helou, Kasliwal, Poznanski, \& Quimby}]{ofek_2012}
Ofek, E., Laher, R., Surace, J., D., L., Sesar, B., Horesh, A., Law, N., van
  Eyken, J., Kulkarni, S., Prince, T., Nugent, P., Sullivan, M., Yaron, Y.,
  Pickles, A., Agueros, M., Bildsten, L., Cenko, S., Gal-Yam, A., Grillmair,
  C., Helou, G., Kasliwal, M., Poznanski, D., \& Quimby, R. 2012, PASP, 1, 1

\bibitem[{Pickles(2010)}]{pickles_2010}
Pickles, A. 2010, in 2010 Space Telescope Science Institute Calibration
  Workshop, edited by S.~Deustua, \& C.~Oliveira (Baltimore: STScI), vol.~1 of
  STSCI, 75

\bibitem[{Pickles \& Depagne(2010)}]{pd_2010}
Pickles, A., \& Depagne, E. 2010, PASP, 122, 1437

\bibitem[{Smith et~al.(2005)Smith, Allam, Tucker, Stute, Rodgers, Stoughton,
  Beers, French, \& McGehee}]{smith_2005}
Smith, J., Allam, S., Tucker, D., Stute, J., Rodgers, C., Stoughton, C., Beers,
  T., French, R., \& McGehee, P. 2005, BAAS, 37, 1379

\bibitem[{Smith et~al.(2002)Smith, Tucker, Kent, Richmond, Fukugita, Ichikawa,
  Ichikawa, A.M., Uomoto, Gunn, Hamabe, Watanabe, Tolea, Henden, Annis, Pier,
  McKay, Brinkman, Chen, Holtzman, Shimasaku, \& York}]{smith_2002}
Smith, J., Tucker, D., Kent, S., Richmond, M., Fukugita, M., Ichikawa, T.,
  Ichikawa, S., A.M., J., Uomoto, A., Gunn, J., Hamabe, M., Watanabe, M.,
  Tolea, A., Henden, A., Annis, J., Pier, J., McKay, T., Brinkman, J., Chen,
  B., Holtzman, J., Shimasaku, K., \& York, D. 2002, AJ, 123, 2121

\bibitem[{Steinbring et~al.(2009)Steinbring, Cuillandre, \&
  Magnier}]{steinbring_2009}
Steinbring, E., Cuillandre, J.-C., \& Magnier, E. 2009, PASP, 121, 295

\bibitem[{Tucker et~al.(2006)Tucker, Kent, Richmond, Annis, Smith, Allam,
  Rodgers, Stute, Adelman-McCarthy, Brinkmann, Doi, Finkbeiner, Fukugita,
  Goldston, Greenway, Gunn, Hogg, Ichikawa, Ivezic, Knapp, Lampeitl, Lee, Lin,
  McKay, Merelli, Munn, Neilsen, Newberg, Richards, Schlegel, Stoughton,
  Uomoto, \& Yanny}]{tucker_2006}
Tucker, D., Kent, S., Richmond, M., Annis, J., Smith, J., Allam, S., Rodgers,
  T., Stute, J., Adelman-McCarthy, J., Brinkmann, J., Doi, M., Finkbeiner, D.,
  Fukugita, M., Goldston, J., Greenway, B., Gunn, J., Hogg, D., Ichikawa,
  S.-I., Ivezic, Z., Knapp, G., Lampeitl, H., Lee, B., Lin, H., McKay, T.,
  Merelli, A., Munn, J., Neilsen, E., Newberg, H., Richards, G., Schlegel, D.,
  Stoughton, C., Uomoto, A., \& Yanny, B. 2006, Astr. Nachr., 327, 821

\bibitem[{Zacharias et~al.(2004)Zacharias, Monet, Levine, Urban, Gaume, \&
  Wycoff}]{zacharias_2004}
Zacharias, N., Monet, D., Levine, S., Urban, S., Gaume, R., \& Wycoff, G. 2004,
  AAS, 205, 4815

\end{thebibliography}

\end{document}